\documentclass[pdflatex,sn-mathphys-num, iicol]{sn-jnl}% Math and Physical Sciences Numbered Reference Style[

\usepackage{graphicx}%
\usepackage{multirow}%
\usepackage{amsmath,amssymb,amsfonts}%
\usepackage{amsthm}%
\usepackage{mathrsfs}%
\usepackage[title]{appendix}%
\usepackage{xcolor}%
\usepackage{textcomp}%
\usepackage{manyfoot}%
\usepackage{booktabs}%
\usepackage{algorithm}%
\usepackage{algorithmicx}%
\usepackage{algpseudocode}%
\usepackage{listings}%
\usepackage{siunitx}
\usepackage{url}

\usepackage{breakurl}
\usepackage[switch]{lineno}
\theoremstyle{thmstyleone}%
\theoremstyle{thmstyletwo}%

\theoremstyle{thmstylethree}%

\begin{document}

\title[Xenon Signal Denoising via Supervised, Semi-Supervised, and Unsupervised Models]{Xenon Signal Denoising via Supervised, Semi-Supervised, and Unsupervised Models}

\author*[1]{\fnm{Grant} \sur{Parker}}\email{parker79@llnl.gov}

\author[1]{\fnm{Jason} \sur{Brodsky}}\email{brodsky3@llnl.gov}
\equalcont{These authors contributed equally to this work.}

\author[1]{\fnm{Indra} \sur{Chakraborty}}\email{chakraborty3@llnl.gov}
\equalcont{These authors contributed equally to this work.}

\affil*[1]{\orgname{Lawrence Livermore National Laboratory}, \city{Livermore}, \postcode{94550}, \state{CA}, \country{USA}}

\abstract{This study presents a denoising algorithm trained using machine learning to improve the energy resolution of a single-phase liquid xenon time projection chamber for neutrinoless double beta decay detection. Supervised, unsupervised, and semi-supervised models are demonstrated to significantly remove noise from simulated measurements while preserving signal information. The supervised model achieves an energy resolution of $<1\%$, while the semi-supervised models achieve energy resolutions of $\sim 1\%$, and the unsupervised model performance is $\sim1.5\%$. This work is evidence that machine learning denoising can improve energy resolution compared to traditional algorithms, even when experimentalists lack perfect \textit{a priori} knowledge of the signals. Such models provide a realistic path toward next-generation sensitivity in $0\nu\beta\beta$ searches.}

\maketitle

\section*{Author ORCID Information}
\noindent Grant Parker: 0000-0003-4957-565X\\
Jason Brodsky: 0000-0002-7498-6461\\
Indra Chakraborty: 0000-0002-0132-8417

\section{Objective}
\label{Objective}

Machine learning (ML) denoising algorithms were developed and evaluated on signals drawn from a simulation of the proposed nEXO neutrinoless double-beta decay ($0\nu\beta\beta$) detector. The denoising methods of this study were evaluated on their ability to reduce the uncertainty on the total charge present in an event, as precision in this measurement has direct impact on the success of this form of particle detector.

\subsection{Background: Neutrinoless Double Beta Decay and the nEXO Experiment}

Observation of $0\nu\beta\beta$ would demonstrate lepton number violation and confirm the Majorana properties of the neutrino, with profound impacts for both the Standard Model of particle physics as well as cosmological models \cite{bariogenesis1,bariogenesis2}. Time projection chambers (TPCs) filled with gaseous or liquid isotopes are one of the primary approaches to searching for $0\nu\beta\beta$  \cite{0nubb_review_2019}, and one example of this TPC strategy is the nEXO \cite{nexo_main} experiment, a xenon-filled TPC enriched with target decay parent $^{136}\text{Xe}$. This study employs nEXO event simulation to train denoising models and ascertain the improvement in energy resolution.

\begin{figure}[h]
\includegraphics[width=.48\textwidth]{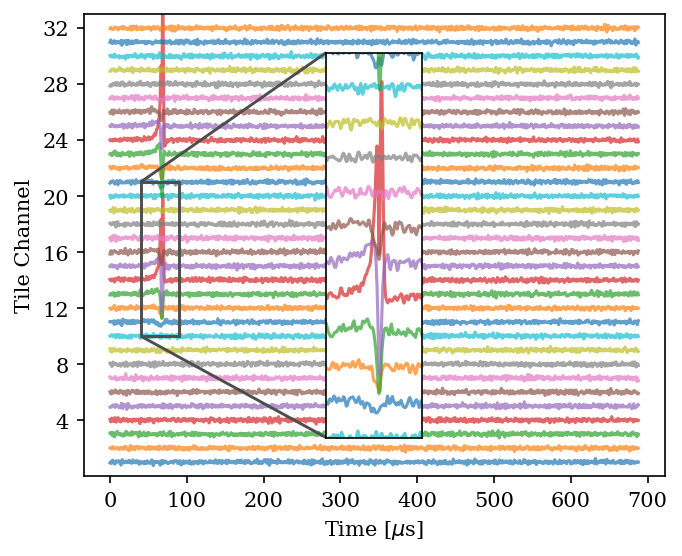}
\caption{\label{fig:waveform_example} Example readout of a single 32-channel tile for a simulated event with added instrumentation noise. Insert: magnified view for portion of signal region}
\end{figure}

The $0\nu\beta\beta$ decay of $^{136}\text{Xe}$ produces two charged $\beta$ particles with total energy equal to the $^{136}\text{Xe}$ decay Q-value of 2458 keV \cite{q_value}. Such $\beta$ particles produce ion tracks along their trajectories, which then drift to the detector charge collectors. The collected charge, illustrated in Fig.~\ref{fig:waveform_example}, in addition to $\beta$-induced scintillation light, is then used to reconstruct the original decay Q-value and infer if the decay was a $0\nu\beta\beta$ event. Electronic noise in charge collection channels adds uncertainty to the measurement of true charge, and by extension, true energy. Therefore, removing charge noise with a denoising algorithm increases the precision of the charge (energy) measurement, which improves overall experiment sensitivity to $0\nu\beta\beta$.

\subsection{Denoising}

Denoising is the process of estimating the ground truth signal contained in a noisy observation. After denoising, further analysis of the signal is both more precise and less complex than attempting to measure quantities directly from the noisy signal.

Supervised denoising models, where the data consists of pairs of true (``clean'') and corrupted (``noisy'') data events, are traditionally deep neural networks (DNNs), most often convolutional neural networks (CNNs) \cite{DnCNN}. A popular supervised framework is the U-Net \cite{unet}: a CNN autoencoder (AE) that includes skip connections. Such models have been successfully applied across multiple noise removal applications \cite{unet_comp1, unet_comp2, unet_comp3}. 

The noiseless measurement data required by supervised models is unavailable from constructed TPCs, as all experimental data is collected with instrumental noise. Simulations based on experiment designs can provide clean-noisy event pairs, yet often have limited accuracy due to incomplete knowledge of the physics and detector response. As such, the performance of a deployed detector may not sufficiently align with simulation. This mismatch can lead to suboptimal supervised denoising performance in experimental applications.

Unsupervised denoising does not require noiseless signals during training. Instead, the model infers across noisy events what the true signal characteristics are. This is analogous to an experimentalist inferring a true signal shape from repeated noisy measurements of an identical signal through averaging. For $0\nu\beta\beta$ detection, repeated measurements of identical events are not possible. Therefore ML methods, rather than simple averaging, are necessary to infer the clean signal characteristics.

Unsupervised learning has two distinct limitations: first, the best-performing methods require exact knowledge of the noise distribution \cite{sure}. In the $0\nu\beta\beta$ context, experiments can characterize noise to excellent precision, as the operation of such experiments ensures there are sufficient noise-only calibration measurements collected between events. Second, unsupervised learning requires more complex loss functions, leading to a loss convergence less rapid and well-behaved than in supervised training. As a result, unsupervised loss functions are more susceptible to converging to model weights in regions where the loss is low yet not optimally-low and the loss function cannot escape. Thus, trained unsupervised learning denoisers typically underperform compared to supervised models with identical mathematical denoising limits \cite{lim1, lim2, lim3, unsure}.

A compromise between the amount of signal-noise knowledge and model performance is semi-supervised learning. The term ``semi-supervised'' is used to describe a wide range of methods. Here, semi-supervised learning pertains to methods that utilize the approximate true signal knowledge of an experimentalist (i.e. an approximately-correct simulation) combined with \textit{post hoc} experimental data that contains the exact signal characteristics in addition to noise. This method can improve performance over a fully-unsupervised model while only placing realistic requirements on the experimentalist to design an approximate signal simulation.

\subsection{Existing Methods and This Article's Contribution}

Noise in the simulated events of this study follows a colored Gaussian profile with a scale of 150 electrons/\SI{0.5}{\micro\second} per timestamp, per channel, modeled after the instrumentation design for the nEXO detector. For comparison, a typical $0\nu\beta\beta$ event in the simulation produces approximately 123K detected electrons, resulting in a relatively large signal-to-noise (SNR) ratio. This high SNR ratio highlights the unique challenge of $0\nu\beta\beta$ denoising compared to the objectives outlined in other denoising literature. Simulated events in this work have a peak-signal-to-noise ratio (PSNR), before any denoising, of 50 dB. By comparison, an example of previous denoising work with medical MRI data utilized images with a PSNR of 32 dB, achieving a PSNR of 38 after denoising \cite{MRIexample}. Hence, this study seeks to further develop existing methods to meet the specific needs of TPC particle detection.

Previous development of ML denoising for $0\nu\beta\beta$ signals is outlined in Ref. \cite{germanium_anderson}, which this work seeks to build upon. In Ref. \cite{germanium_anderson}, a supervised AE trains on simulated 1D signal pulses for a crystal $^{76}\text{Ge}$ detector, where the target is a ``clean'' signal event with no noise and the input is the target with added real detector noise and Gaussian white noise. Additionally, a model trained on approximations of the simulation tested the transfer of denoising models trained without perfect knowledge of the signals (i.e. a ``semi-supervised'' model). Lastly, an unsupervised model was trained through a modification of the Noise2Noise \cite{n2n} methodology, where the only input was the noisy signal event. Notably, the Noise2Noise as originally described in Ref. \cite{n2n} utilizes pairs of signals with independent noise, and the method in Ref. \cite{germanium_anderson} is more similar to successors of Noise2Noise that use unpaired noisy inputs, such as Noisier2Noise \cite{noisier2n} and Recorrupted2Recorrupted \cite{r2r}.

This work presents models in a similar format to Ref. \cite{germanium_anderson} with modifications and improvements: First, models in this work denoise multichannel events, in which the same event is observed across several sensors. This has the result of increasing the dimension of the input from \num{4096} (1 channel $\times$ \num{4096} timestamps) in Ref. \cite{germanium_anderson} to \num{44064} (32 channels $\times$ \num{1377} timestamps) in this work. To maximize effectiveness, rather than denoise individual 1D channels, the denoiser must learn correlations within the signal across both time and sensor channels. Second, compared to Ref. \cite{germanium_anderson}, this work's denoising achieves a greater improvement in energy resolution relative to a trapezoidal filter. Third, this work demonstrates a semi-supervised approach that significantly improves energy resolution even when trained on highly inaccurate simulations. Ref. \cite{germanium_anderson} reported no significant energy resolution improvement on experimental data, and attributes this to differences between the training and experimental data. This study demonstrates that even large simulation-experiment discrepancies can be managed effectively with the semi-supervised approach.

\subsection{Outline}
The contents of this article are as follows: 
In Section \ref{Data}, the data employed for training and validation is described, along with the preparation for each model type. Section \ref{Benchmarking} defines a physics-motivated parameter related to the energy resolution that tracks model performance during training, as well as the mathematical limit of performance that a given model could achieve. Section \ref{sec:supervised} reviews the architecture of the supervised U-Net denoising model and its loss function. Section \ref{sec:method-unsupervised} then presents how the model of Section \ref{sec:supervised} can be trained solely on noisy data in a completely unsupervised mode with an alternative loss function. Lastly, the semi-supervised mode is presented in Section \ref{sec:semi_supervised}, where the first stage trains on distorted events without noise in the supervised mode, and the following stage trains on undistorted events with noise in the unsupervised mode. Results are presented in Section \ref{Results}, with discussion following in Section \ref{Conclusions}.

\section{Data}
\label{Data}

The dataset used for model training and validation consists of $\sim$83M simulated drift charge detection events from a single-phase xenon TPC, produced via methods described in Refs.\cite{simulation, nexo_main, nexo_sens_2}. The simulated signals are charge currents measured on multi-channel electrode tiles, where an example of a simulated event on a single tile is illustrated in Fig.~\ref{fig:waveform_example}. 

Electrons collected by an electrode produce a large pulse of positive current on that electrode. In addition, all electrodes near the drifting electron cloud experience induction, producing a bipolar current pulse with zero net current. This induction spreads information about the drift charge across multiple channels and time points, increasing the number of noisy samples overlapping with signal information and thus increasing the value of denoising.

Target decays are either contained locally at a single ``site'' or span the fiducial volume as a multi-site event, and as such, charges are often collected by multiple tiles \cite{nexo_main}. For simplicity, models in this study are trained on simulated single-tile events, which allows for the denoising of multi-tile events through compiling individual denoised tile results. Hence, for this study, each simulated event contains 32 channels corresponding to the 32 sensing electrodes of a single tile within the TPC, and 1377 time increments, each a \qty{0.5}{\micro\second} sample from a digitizer, representing the maximum possible drift time in the simulated TPC. Therefore each signal has a total of \num{44064} dimensions per event.

In addition to the drift charge simulation, the events include simulation of energetic particle behavior using Geant4 \cite{geant}. Events include a mixture of $0\nu\beta\beta$ and gamma rays, which produce a mixture of single- and multi-site background ionization in the TPC. All events used for training, including $0\nu\beta\beta$, were simulated with a uniform range of kinetic energy between \SIrange[]{700}{4000}{\kilo\electronvolt}. Although this energy is not physical for $0\nu\beta\beta$, it ensures the denoiser is trained without a bias towards any particular energy. Model performances are evaluated on a holdout ensemble of 69K $0\nu\beta\beta$ events, separate from the training and validation events, simulated only at the physical Q-value to determine the fractional energy resolution, $\sigma_E/E$.

For each event the true (noiseless) signal, $y$, is recorded, as well as $x$, the true signal with electronic noise added. The power spectrum density (PSD) of the noise is determined from measurements of prototype TPC electronics, and follows a $1/f$ profile, where $f$ is the frequency \cite{noise1, noise2}. In addition, $x$ is quantized to reflect the behavior of a 12-bit digitizer, which introduces additional noise. Both $x$ and $y$ are represented in arbitrary digitizer units, such that 9 digitizer units are equivalent to one electron per \qty{0.5}{\micro\second} sample. For supervised training, paired $x$ and $y$ are employed; during unsupervised training, only $x$ is contained in the input. The semi-supervised mode is a two-stage process where for a specified proportion of data, both $x$ and $y$ are available, then only $x$ is utilized for the remainder of training.

\section{Benchmarking}
\label{Benchmarking}

For standardized evaluation of all considered denoising models, metrics relevant to the physics objectives of TPC experiments are employed. The primary objective is sufficient energy resolution to discriminate detector events, and for the purposes of this study, improved energy resolution is pursued via improved charge resolution. The integral of the digitized current readout (across all timestamps and sensor channels) within an event is proportional to the total charge deposit. Therefore, the primary error metric of this study is the following:

\begin{gather}
    \eta_Q(y)  = \frac{\sum_{c,t} \hat{y} - \sum_{c,t} y }{\sum_{c,t} y } ,
\end{gather}

\noindent where $c$ and $t$ are the channel and timestamp indices, respectively, and $\hat{y}$ is the model prediction. Hence, $\eta$ is the fractional integral error, defined as the relative difference between the integral of the denoised prediction, $\hat{y}$, to that of the noiseless signal, $y$. This relative error is identical to the relative error of the charge estimate:

\begin{gather}
\label{eta_q}
    \eta_Q = \frac{\hat{N}_Q - \tilde{N}_Q }{\tilde{N}_Q} ,
\end{gather}

\noindent such that $N_Q$ is the total event drift charge, given $ N_Q = \alpha \sum_{c,t} y$, with $\alpha$ the scale factor introduced by the digitizer.

\subsection{Relationship of \texorpdfstring{$\eta$}{eta} to energy resolution}

A single-phase TPC, as simulated here, measures both scintillation photons and drift electrons to estimate the energy of an event. Total photon and electron production is proportional to the energy of the particle interaction \cite{nexo_main}:
\begin{gather}
    E = (\tilde{N}_P + \tilde{N}_Q)W_q \, ,
\end{gather}

\noindent where $W_q$ is the averaged energy of producing a single photon or charge unit. The fractional energy resolution at $0\nu\beta\beta$ decay energy is then

\begin{equation}
\frac{\sigma_E}{E} \bigg|_{0\nu\beta\beta} = \frac{W_q}{E}  \sqrt{ \sigma_P^2 + \sigma_Q^2 + 2\rho \sigma_P \sigma_Q + \sigma_r^2 + \sigma_\text{Xe}^2} \,,
\end{equation}

\noindent where $\sigma_P^2$ and $\sigma_Q^2$ are the photon and charge variances, respectively, $\rho$ is the charge-photon correlation coefficient, $\sigma_r^2$ the contribution from recombination, and $\sigma_\text{Xe}^2 = f_\text{Xe} \langle n \rangle$ is the term that accounts for non-recombination fluctuations in the total quanta. For direct comparison to the results in Ref. \cite{nexo_main}, this study assumes the photon-charge anticorrelation term perfectly cancels with recombination, that the Fano-like term is negligible, and that electron drift loss from volume impurities is negligible, giving our final prescription for the fractional energy resolution:

\begin{equation}
\frac{\sigma_E}{E} \bigg|_{0\nu\beta\beta} \approx \frac{ \sqrt{ \sigma_P^2 + \sigma_Q^2 }}{\langle n \rangle} \,, 
\end{equation}

\noindent with $\langle n \rangle$ the total expected photon-electron quanta for an event, given by $E = \langle n \rangle W_q$. Therefore, improvements in charge resolution $\sigma_Q$ directly improve energy resolution. 

The fractional energy resolution determines the overall performance of a model, but cannot be efficiently determined for \textit{in situ} monitoring of model training. Given the fractional charge error of an event provided by Eq. \ref{eta_q}, the charge error in electron units is determined by $\sigma_Q =  Q  \langle \eta_Q \rangle $, where $Q$ is the expected number of electrons for a $0\nu\beta\beta$ event. While the final model value of $\langle \eta_Q \rangle$ is determined from the independent holdout dataset after training, for a batch of validation events in an epoch step, the rapid calculation of $\langle \eta_Q \rangle_\text{batch}$ allows for a physics metric that tracks the approximate improvements in $\sigma_Q/Q$ and $\sigma_E / E$ made by a model over the course of training. Final model evaluations of $\sigma_Q/Q$ and $\sigma_E/E$ are presented in Section \ref{Results}, assuming $\sigma_P \approx 1319.5$ and $\langle n \rangle = $\num{213747} following Refs.\cite{nexo_main, simulation, sipm}.

\subsection{Optimal filter}
For data with signals and noise that are stationary, or have time-independent statistical characteristics, a bound on the best possible denoising algorithm is set by the Wiener filter \cite{wiener_original}: a linear minimum mean squared error (MMSE) estimator of the unknown amplitude of a known signal template affected by noise with a known power spectrum. Under these assumptions, the output of the Wiener filter is the optimal reconstruction of the true event amplitude. 

This ``optimal filter'' output, $\hat{y}$, of a noisy event, $x$, is determined by the following: assuming the shape of the true signal $y$ is known, $\hat{y}$ may be written as

\begin{equation}
\hat{y}(t) = Ay(t) + n(t),
\end{equation}

\noindent where $A$ is an unknown amplitude and $n(t)$ is the time-dependent additive noise. As such, the optimal filter estimate for the noisy signal is

\begin{equation}
\hat{y}_{\text{optimal}}(t) = \hat{A}y(t),
\end{equation}

\noindent where $\hat{A}$ is the optimal linear estimator for $A$. If $\hat{Y}_k = \text{rFFT}\{\hat{y}\}_k$ and $Y_k = \text{rFFT}\{y\}_k$ are the real-valued Fourier coefficients across frequencies $f_k$ and the noise has amplitude spectral density (ASD) $p_k$, then $\hat{A}$ is calculated via

\begin{equation}
\hat{A} = \frac{\sum_k \hat{Y}'_k Y'^{*}_k}{\sum_k |Y'_k|^2} \quad, 
\end{equation}

\noindent where $\hat{Y}'_k = \hat{Y}_k / p_k$ and $Y'_k = Y_k / p_k$ are the whitened Fourier coefficients.

The ML denoising models studied here have \emph{less} knowledge than assumed by this optimal filter. In the optimal filter, only the signal amplitude is uncertain while the signal shape of each event is known. An ML denoiser confronts uncertainty in both the signal amplitude and shape of each event. While an ML denoiser with supervised training will observe the ground truth signal shapes of the training events, it does not not know \textit{a priori} the shape of a new, previously unseen, event upon evaluation, and must infer the shape from the noisy input. An unsupervised denoiser has even more uncertainty as to the signal shape, as it has never observed a noiseless version of any signal. As such, the optimal filter places a bound on the performance of any experimentally-realistic ML denoiser. This study examines how close ML algorithms can approach the optimal filter limit without the expectation of matching that limit.

\section{Supervised Models}
\label{sec:supervised}

The supervised model of this study is a U-Net-style autoencoder (described in Table \ref{tab:model}) trained to learn a denoising function $g(x) = \hat{y}$. This denoising function, $g$, was trained to minimize the Smooth L1 Loss \cite{smoothl1} between the model prediction $\hat{y}$ and the true signal $y$: 

\begin{gather}
\mathcal{L}_\text{SL1}(\hat{x},x) = \begin{cases} 
\sum\limits_{i=0}^{N} \frac{(y_i-\hat{y}_i)^2} {2N\beta}, & \text{if}\,\,\,|y_i-\hat{y}_i|<\beta \\
|y_i-\hat{y}_i|-0.5\beta, & \text{otherwise}
\end{cases}
\end{gather}

\noindent where $\beta$ is a hyperparameter.
L1 loss is preferred in this application for its robustness to event outliers, stability at small loss values, and simultaneous minimization of the signal integral employed for total charge estimation as described in Section \ref{Benchmarking}.

\begin{table}[ht]
\caption{Specifications for U-Net architecture employed for all models, including layer output dimensions. All models in this study have base channels (BC) = 512.}
\label{tab:model}
\centering
\renewcommand{\arraystretch}{1.3} % 1.0 is default
\begin{tabular}{ ccc }
\toprule
& Layer &   Output\\
\midrule
\multirow{5}{*}{\rotatebox{90}{Encoder}} & Input & $32 \times 1377$ \\
& (Strided) Convolution \#1 & $\text{BC} \times 689$ \\
& (Strided) Convolution \#2 & $\text{BC} \times 345$ \\
& \rotatebox{90}{...} & \\
& (Strided) Convolution \#10 & $\text{BC} \times 2$ \\
\midrule
\\
& Bottleneck Convolution ($\times 24) $ & $\text{BC} \times 2$ \\
\\
\midrule
\multirow{11}{*}{\rotatebox{90}{Decoder}}& Concatenate Convolution \#10 & $(2 \times \text{BC}) \times 2$ \\
& Upsample \#1 & $\text{BC} \times 4$ \\
& Convolution \#11 & $\text{BC} \times 4$ \\
& Concatenate Convolution \#9 & $(2 \times \text{BC}) \times 3$ \\
& Upsample \#2 & $\text{BC} \times 6$ \\
& Convolution \#12 & $\text{BC} \times 6$ \\
& \rotatebox{90}{...} & \\
& Concatenate Convolution \#1 & $(2 \times \text{BC}) \times 689$ \\
& Upsample \#10 & $\text{BC} \times 1378$ \\
& Convolution \#20 & $\text{BC} \times 1378$ \\
& Final Convolution (No Activation) & $32 \times 1377$
\\
\bottomrule
\end{tabular}
\end{table}

A traditional autoencoder with no skip layers, in addition to a U-Net, was investigated in this study. The traditional autoencoder has the notable feature of dropping information during downsampling layers, with the bottleneck storing significantly less information than the input. As the clean signals contain less entropy than the noisy ones, this approach is appealing: the loss of noise information while preserving signal information would, in principle, lead to an effective denoiser. However, during testing, it was observed that small bottlenecks lead to a loss of signal information in addition to noise information, causing poor denoiser performance. This motivated this study's employment of a U-Net architecture, which transfers the complete information across the bottleneck.

The model architecture employed across supervision levels in this study is a symmetric U-Net constructed with one-dimensional convolutions in the time dimension. The signal channel dimension is left fully-connected rather than utilizing 2D convolutions, as the charge collection tile electrodes do not have a pixel-like adjacency relationship. The number of layers is selected to ensure the receptive field of the network is capable of combining information across samples at the earliest and latest times in the measurement.

\begin{figure}[t]
\includegraphics[width=.48\textwidth]{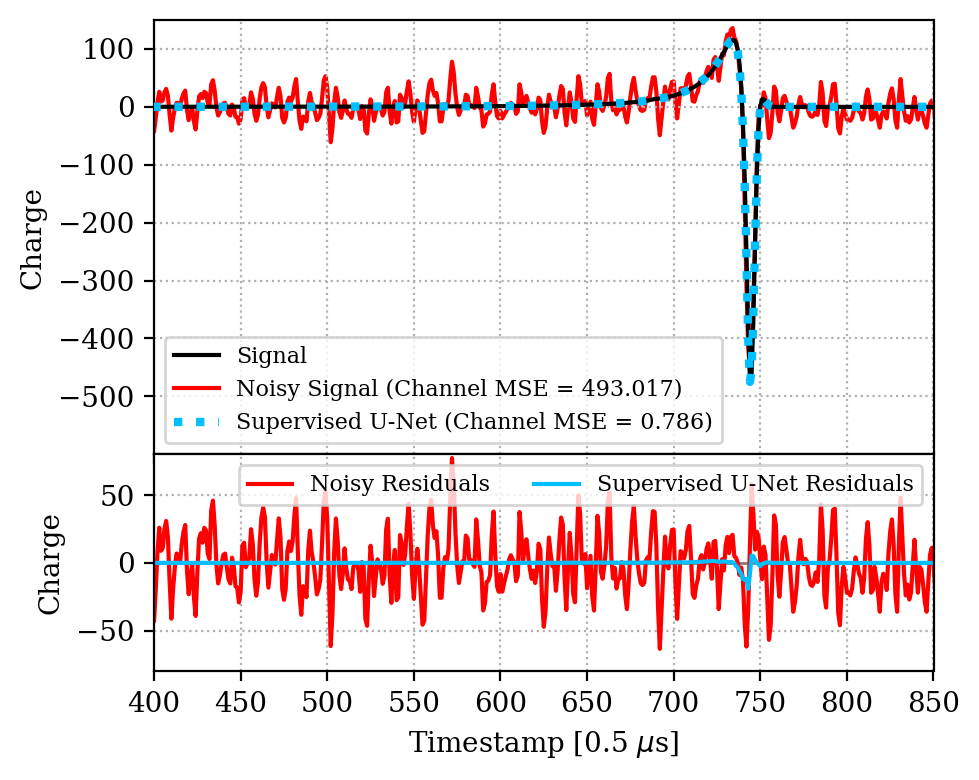}
\caption{\label{fig:channel_example} Simulated charge readout for a single tile channel. (\textit{Upper}) Model-predicted (blue) charge readout for noisy signal (red) and true signal (black) simulation data for one tile channel. (\textit{Lower}) Residuals for noisy (red) and model prediction (blue) }
\end{figure}

Sampled from the simulated data described in Section \ref{Data}, $\sim$83M pairs of clean-noisy events are used for training, with 200K events reserved for validation. The model was trained for 300K steps with a batch size of 256, representing approximately a single pass through the training set. This training required 48 hours on a node of four Nvidia H100 GPUs. An example of the output from this supervised model for a single event channel is provided in Fig.~\ref{fig:channel_example}.

\section{Unsupervised Models}
\label{sec:method-unsupervised}

For the unsupervised model of this study, an identical neural network architecture as in Section \ref{sec:supervised} is utilized, but with a larger amount of base channels (greater capacity) and a loss function that does not require access to $y$, the ground truth clean signal. For the loss, Stein's Unbiased Risk Estimator (SURE) \cite{sure} is substituted, which estimates the ability of a function to remove noise without knowledge of the true signal. Minimizing the SURE loss thus trains an optimal denoiser without need for preexisting ground truth signal knowledge. The SURE loss is given by the following function for the case of Gaussian white noise with variance $\sigma^2$ \cite{sureloss}:

\begin{gather}
    \mathcal{L}_\text{SURE} = \text{MSE}(x,\hat{y}) + 2\sigma^2 \text{div}[f(x)]
\end{gather}

\noindent where $\text{MSE}(x,\hat{y})$ is the mean square error, $f(x)$ is the unsupervised denoising function, and $\text{div}$ is the divergence operator. Note there is no dependence on true signal $y$, only noisy input $x$ and prediction $\hat{y}$.

As the simulated electronic noise in this study is not white but has a colored spectrum with covariance matrix $\Sigma$, a modified SURE loss is required\cite{unsure}:

\begin{gather}
    \mathcal{L}_{\text{UNSURE}} = \text{MSE}(x,\hat{y}) - n \text{Tr}[\Sigma] + 2 \text{Tr}\bigg[\Sigma \frac {\partial f}{\partial x}(x)\bigg] 
\end{gather}

\noindent where Tr is the trace operator, $\frac {\partial f}{\partial x}(x)$ is the Jacobian, and $n$ is the dimension of $x$.

The first term of this loss is the reconstruction term, evaluating whether the prediction $\hat{y}$ resembles the noisy input $x$. This term independently pulls the prediction to replicate the input; a function trained to minimize only this first term would converge to the identity function. The second term does not depend on $f$ and therefore is not required for training. However, the term is necessary to employ the SURE calculation as a risk estimate as opposed to its use as a loss for ML training. The third term is the divergence of $f$, colored by the covariance of the noise. This term pushes the function $f$ to be contractive: when provided a larger input, e.g. from a positive noise fluctuation, the function should pull the input negatively, countering the noise fluctuation. This term independently would prompt the model to remove true signal; only in combination with the first term does the third term lead to denoising that preserves the ground truth.

During training, calculating $\frac {\partial f}{\partial x}(x)$ exactly is computationally unfeasible due to $y$ having dimension $32\times1377 = \num{44064}$. With $\mathcal{O}(10\text{k})$ divergence terms to be calculated for each of the $\mathcal{O}(10\text{M})$ training events, the computational requirements would rapidly inflate above those necessary for the supervised training.

This study resolves the complication via the MC-SURE method \cite{mcsure}. Partial derivatives are approximated through a finite-difference calculation, and the trace across $\sim$44K dimensions is approximated with a Hutchinson trace estimator sampling random vectors in the 44K-dimension space. The full application is given by:

\begin{gather}
2 \text{Tr}\bigg[\Sigma \frac {\partial f}{\partial x}(x)\bigg] \approx \frac{2}{\epsilon k} \sum_{i=1}^{k} z_i^T[f(x+ \epsilon z_i) - f(x)] .
\end{gather}

\noindent Here, $z_i$ are $k$ independent draws from the colored noise distribution and $\epsilon$ is a small scaling factor for the finite-difference approximation.

This method introduces two hyperparameters, $k$ and $\epsilon$, that must be tuned per each use case. Larger values of $k$ reduce the error on the Hutchinson trace estimator, as in this scenario, larger $k$ translates to a larger sample of the 44K dimensions for use in the approximation of $\frac {\partial f}{\partial x}(x)$. However, larger $k$ also increases computing time, reducing the number of training steps per unit time. The standard deviation of $k$ was evaluated during preliminary training and was found to be small compared to the event-to-event variation in the overall loss. As a result, this study implemented $k$ = 2, sufficient for monitoring $\sigma_k$ yet prioritizing higher amounts of training steps. Smaller values of $\epsilon$ reduce errors from the finite-difference approximation, yet an extremely small $\epsilon$ induces floating-point errors in $f$. Therefore, $\epsilon$ = 0.1 was determined through testing for the final unsupervised training.

Due to the quantization of the noisy signals described in Section \ref{Data}, the noise profile of the training data does not completely fulfill the assumptions in the SURE loss derivation. To account for this effect, identical quantization has been applied to the injected noise probe, $z$, with the other MC-SURE loss function terms remaining unaltered. With this modification, the training was observed to proceed as designed, reducing both the SURE loss and the integral error $\eta_Q$, for approximately the first 37\% of training events. Beyond this threshold, an increase in the variance of the $k$ independent noise probes of the divergence term was noted. This increase in probe variance corresponds with a rapid degradation of the performance of the denoiser as measured by the integral error. It is hypothesized that at this threshold, the model identifies integer outputs minimizing the reconstruction term better than float outputs, as the noisy input employed in the reconstruction term comparison is integer-valued. This ``snapping'' to integers behavior has large divergences at the cusps between integer values, and it is hypothesized that the noise probes inconsistently measure these cusps. As a result, the final models of this study have training stopped early when the variance between the $k$ noise probes transitions from decreasing across steps to increasing.

To evaluate if the finite-difference approximation exacerbates the effects of noise quantization, an alternative approximation for the second term of the MC-SURE loss function was tested:

\begin{gather}
2 \text{Tr}\bigg[\Sigma \frac {\partial f}{\partial x}(x)\bigg] \approx \frac{2}{k} \sum_{i=1}^{k} z_i^T J[f(x)].
\end{gather}

\noindent where $J[f(x)]$ is the Jacobian of $f$. To avoid the full calculation of the Jacobian, the vector-Jacobian product $z^T J$ is computed through the PyTorch autograd capability. This method performed sufficiently similar to the finite-difference method and was not able to eliminate the increase in the variance between the noise probes. Hence, the finite-difference approximation was selected for the final model training in the unsupervised scenario.

SURE was selected as the unsupervised loss function of this study after first investigating Recorrupted2Recorrupted (R2R) \cite{r2r}. R2R is the most similar method in the literature to the modification of Noise2Noise used in Ref. \cite{germanium_anderson}. However, R2R applied to the signals in this study could not achieve $\sigma_Q/Q < 2.8\%$. Following the literature, it was noted that SURE loss optimizes a similar objective to R2R \cite{gen-r2r}, and therefore the SURE method was attempted and observed to have superior performance. This superior performance can be attributed to two effects: first, R2R optimizes denoising for the \emph{recorrupted} signals, i.e. in the case of additional noise injection, while SURE optimizes denoising for the original measurement noise. Second, R2R is exposed to gradient noise due to the random noise probe used to recorrupt the input and target. MC-SURE is also exposed to gradient noise due to a random noise probe, yet this noise only appears in the divergence term, not the comparatively larger reconstruction term. The authors of this study hypothesize the differences are significant only when the denoiser is attempting to achieve the very low level of residual noise targeted in this study; on inherently noisier signals these effects may be less significant, leading to little difference in the performance of SURE versus R2R, as observed in Ref. \cite{gen-r2r}.

\section{Semi-Supervised Models}
\label{sec:semi_supervised}

Semi-supervised learning in this study aims to replicate the common scenario of an experimentalist leveraging the limited knowledge available: approximate knowledge (simulation) of the ground truth signals, and noisy experiment measurements containing the exact ground truth signals. This method was pursued after observing that the fully-unsupervised method of Section \ref{sec:method-unsupervised} did not achieve as low a value of the SURE loss as the supervised model had achieved. It was then hypothesized that the training procedure of the unsupervised model would converge the model weights to a region of lower loss, yet fail to escape the region to further minimize the loss. 

However, while incomplete, the hypothetical experimenter's knowledge is sufficient to achieve significant noise removal via the following procedure: in Stage I, a supervised denoiser is trained on the simulation, as described in Section \ref{sec:supervised}. While the simulation only approximates the ground truth, the model proceeds to learn broader signal features and direct weight gradients in the direction of the greater loss minima. In Stage II, model training with the weights from Stage I resumes, yet now the training follows the unsupervised methods described in Section \ref{sec:method-unsupervised}, correcting output of Stage I to match the signals present in unlabeled (i.e. real detector) data instead of the flawed simulation. It is theorized that pre-training the model in a supervised mode on distorted data allows the unsupervised mode to find a different, better region with lower local minima than that found by solely-unsupervised model, improving denoising performance.

\begin{figure}[t]
\includegraphics[width=.48\textwidth]{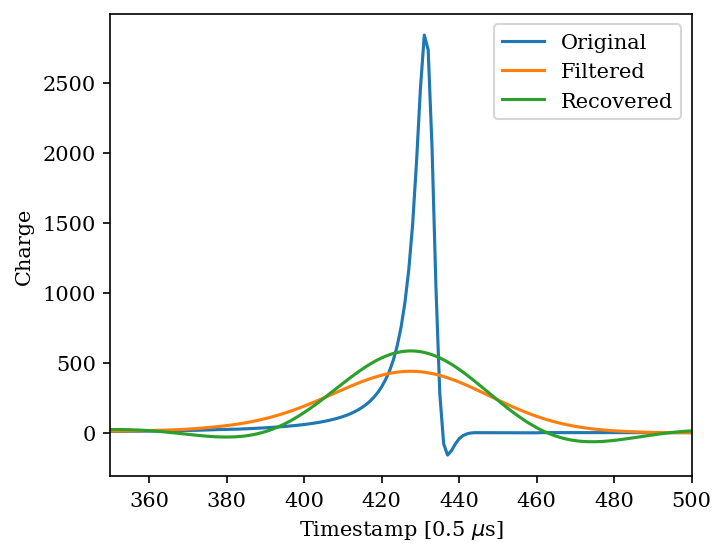}
\caption{\label{fig:data_loss} Example of the Gaussian filter applied to generate the 55\%-distortion training set. The recovered signal is an attempt to invert the distortion, demonstrating that the distortion represents significant loss of the original information}
\end{figure}

In this study, all signals are the result of a simulation. To model the imperfect knowledge of an experimentalist, this study divides the original simulation into two sets: the \textbf{exact set} is the unmodified output of the simulation, as described in Section \ref{Data}, while the \textbf{distorted set} is created through the application of a low-pass filter to the signal events, deliberately degrading the information in those signals. Stage I of the semi-supervised models in this study is trained on pairs of noisy and noiseless events from the distorted set. Stage II trains on the exact set, but only uses the noisy version of those events and not the noiseless version. Noiseless, un-distorted events are not used in either stage.

Four distorted sets were created for this study, each with a different degree of alteration from the original signals, to evaluate the effect of better or worse approximation. Each distortion was created by applying a Gaussian filter to the signal (in the time dimension only) with a width specific to that set. We measure the loss of signal shape information by inverting the Gaussian kernel to attempt to recover the original signal. We calculate the information loss of the filter as:

\begin{gather}
D = \frac{\mathbb{E}[(y_{\text{recovered}} - y_{\text{original}})^2]}{\mathbb{E}[y_{\text{original}}^2]} 
\end{gather}

\noindent The four sets have $D = 0.1\%, 10\%, 30\%,$ and $55\%$. An example of the filter that causes 55\% data distortion is shown in Fig.~\ref{fig:data_loss}.

The approach of using a low-pass filter was chosen as a proxy for instances in which a simulation may capture coarser features of instrument signals, yet miss finer details. The most distorted set represents a significantly more coarse simulation ($\sim20\%$ charge integral error) than is likely to occur in particle physics simulations employed for statistical inference. Future work could examine denoiser performance under variations in the imperfect simulation method, such as altering simulation physics parameters.

\begin{figure}[t]
\includegraphics[width=.48\textwidth]{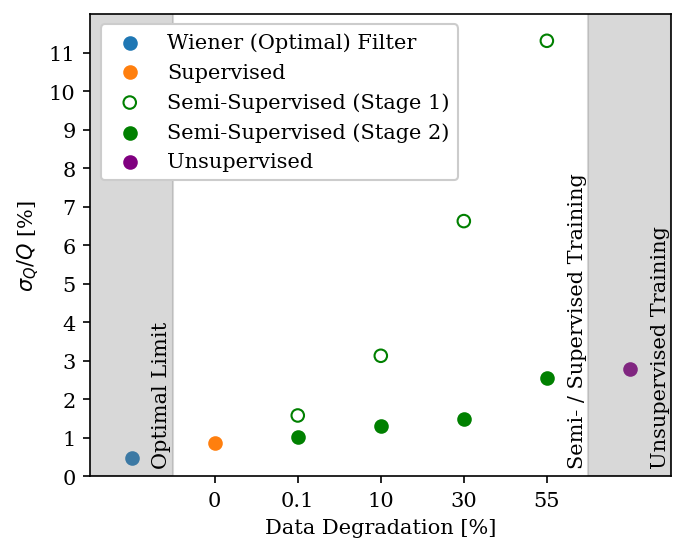}
\caption{\label{fig:charge_vs_degradation} Charge resolutions across models of various data degradations}
\end{figure}

\subsection{Non-ML Denoising}

For the purpose of comparison to the ML denoisers, a selection of conventional denoising methods was investigated. Such selection was informed by Ref. \cite{germanium_anderson} in addition to prior denoising research conducted within the nEXO experiment. One of the selected denoiser methods was Savitzky-Golay filtering \cite{savitz}; in this study, window length 3-151 and polynomial order 1-11 were tested. The best-performing filter had a window length of 5, polynomial order 1. Notably, similar performance was observed across a range of these parameters. Also investigated was wavelet-based denoising with Visushrink \cite{visu}. As with the Savitzky-Golay filter, a parameter optimization was performed over wavelet classes as well as the number of wavelet decomposition levels. The best performance was observed with the Sym12 wavelet and decomposition to eight levels.

Following prior research within nEXO, a trapezoidal filter was employed to estimate the integrals of charge signals. This filter produces an estimate of the integral yet maintains key differences from the other conventional methods discussed here. For instance, the trapezoidal filter does not produce a time series estimate of the clean signal. Additionally, compared to the other methods of this section, trapezoidal filtering integrates a specific window within the signal, excluding noise far from the signal pulse from the integral. While the performance of the trapezoidal filter was measured independently, applying the filter to the outputs of the Savitzky-Golay and Visushrink filters was tested, yet no significant improvements to the integral estimate were observed. The best-performing trapezoidal filter used a rise time of 11 samples, a flat time of 100 samples, and kept the top five adjacent sensor channels in each orientation.

\section{Results}
\label{Results}

\begin{figure}[t]
\includegraphics[width=.48\textwidth]{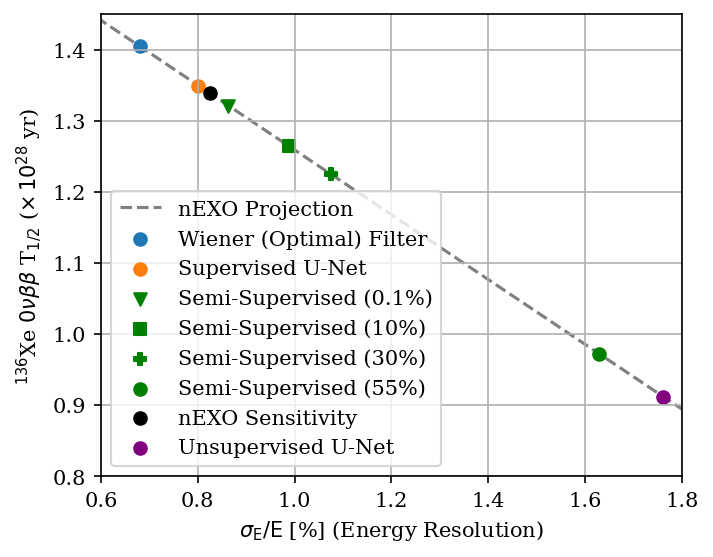}
\caption{\label{fig:halflife_vs_resolution} Projected 90\% median sensitivity as a function of energy resolution, from Ref. \cite{nexo_main}}
\end{figure}

The final energy resolution achieved by each model is determined from a holdout set of events, separate from the training and validation sets, with energy equal to $0\nu\beta\beta$ Q-value, with 69K events total. Table \ref{tab:results} lists the fractional charge resolution $\sigma_Q/Q$ and the corresponding fractional energy resolution, $\sigma_E/E$, of the denoising methods tested in this study. The final $\sigma_Q/Q$ of each ML method as a function of data distortion is presented by Fig.~\ref{fig:charge_vs_degradation}, with the Wiener filter results shown for baseline comparison. Although such filter is impossible to implement in a real experiment, the result is a bound on the best possible model performance. The translation of a given model's charge resolution to energy resolution and half-life sensitivity from the projection derived by Ref. \cite{nexo_main} is displayed in Fig.~\ref{fig:halflife_vs_resolution}.

Among the ML methods, there is a clear trend: the more information about the ground truth signal available during training, the better the resulting $\sigma_Q/Q$. This result is not a mathematical foregone conclusion, as the fully unsupervised method, in theory, has sufficient information from noisy signals to learn the optimal denoising function. However, in practice, the difficulty of training the unsupervised method results in worse performance.

Comparing ML methods to the conventional methods, the ML models return superior performance. The trapezoidal filter significantly improved $\sigma_Q/Q$ from baseline noisy signals, but did not reach the performance of even the fully unsupervised ML denoiser. The Visushrink method only marginally improved $\sigma_Q/Q$, and the Savitzky-Golay filter offered no improvement. The relative success of the trapezoidal filter may be attributed to its ability to completely reject all noise, including low-frequency noise, in time regions distant from the signal portion---a feature that neither Visushrink nor the Savitzky-Golay filter possesses. Combining the trapezoidal filter with Visushrink produced a small improvement to trapezoidal filter alone.

Also listed are the $\sigma_Q/Q$ and $\sigma_E/E$ values reported in Ref. \cite{nexo_main}, the most recent release of the nEXO experiment's projected performance. Such values are reported to demonstrate the ability of the denoising algorithms in this study to meet the requirements of a nEXO-like detector.

To understand the gap in performance between the unsupervised and supervised models, the SURE loss was evaluated on both models after training. The supervised model had $\texttt{SURE}=529.4+0.7$, where the two terms reported here are the reconstruction and divergence terms respectively. The unsupervised model had $\texttt{SURE}=531.4+3.4$. Such values indicate that the unsupervised model did not achieve the minimum possible SURE loss during training, despite the optimization target. Two causes of this failure to minimize SURE are hypothesized: gradient noise in the MC-SURE procedure and the abundance of local minima in the SURE loss space. Either effect or their combination may have forced the training of the unsupervised model to plateau without minimizing to the lower SURE value that the supervised model achieved.

\setlength{\tabcolsep}{7pt}
\begin{center}
\begin{table}[bt]
\caption{Final training performance this study's models, with conventional method performances as comparison}
\label{tab:results}
\centering
\begin{tabular}{ cccc }
\\
\toprule
Model/Ref.                              & $\sigma_Q/Q$  &   $\sigma_E/E$ \\
\midrule
\\[1pt]

Wiener (Optimal) Filter\\(Ref. \cite{wiener_original})   &    0.477\%    &   0.680\%     \\[6pt]
Supervised U-Net \\ (This Work)                         &    0.859\%    &   0.802\%     \\[6pt]
Semi-Supervised U-Net, 0.1\% \\ (This Work)             &    1.014\%    &   0.863\%     \\[6pt]
Semi-Supervised U-Net, 10\% \\ (This Work)              &    1.299\%    &   0.989\%     \\[6pt]
Semi-Supervised U-Net, 30\% \\ (This Work)              &    1.479\%    &   1.075\%     \\[6pt]
Semi-Supervised U-Net, 55\% \\ (This Work)              &    2.536\%    &   1.629\%     \\[6pt]
Unsupervised U-Net \\ (This Work)                       &    2.773\%    &   1.761\%     \\
\\[1pt]
\midrule
\\[1pt]
Trapezoidal Filter                                  &    4.5\%      &   2.746\%     \\[6pt]
Visushrink Wavelet Denoising                        &    12.6\%     &   7.519\%     \\[6pt]
Trapezoidal Filter + Visushrink                     &    4.0\%      &   2.457\%     \\[6pt]
Savitzky-Golay Filter (w=5, p=1)                    &    13.8\%     &   8.230\%     \\[6pt]
No Denoising                                        &    13.8\%     &   8.230\%     \\[6pt]
\midrule
Assumptions in\\nEXO Sensitivity (Ref. \cite{nexo_main})             &    0.921\%    &   0.826\%     \\[6pt]
\bottomrule
\end{tabular}
\end{table}
\end{center}

\section{Conclusions}
\label{Conclusions}

The success of semi-supervised denoising for particle detection signals is directly applicable to the real-world needs of experiments. While obtaining an exact match between simulated signals and experiments is not realistic for any experiment, an approximate match is commonly achieved through simulation based in first-principles physics. Experimentalists should be encouraged by the result here demonstrating that approximate knowledge of the signal combined with unlabeled experimental data can produce a denoiser with nearly the same performance as a fully-supervised training, which significantly exceeds conventional methods.

The challenges of training a fully unsupervised model is a rich topic of interest for future ML research. While semi-supervised training is practical, fully unsupervised training is the optimal scenario. More broadly, further understanding the gap between the theoretical promise of the SURE loss and the practical result of training with the MC-SURE method will, we expect, increase understanding of how to further improve all forms of ML denoising.

\section{Data Availability}

Due to the large volume of data, training datasets generated and analyzed for this study are available upon reasonable request to the corresponding author.

\section{Code Availability}

Model architecture and evaluation code will be available upon reasonable request to the corresponding author.

\section{Acknowledgments}

Lawrence Livermore National Laboratory is operated by Lawrence Livermore National Security, LLC, for the U.S. Department of Energy, National Nuclear Security Administration under Contract DE-AC52-07NA27344. LLNL-JRNL-2016336

\bibliography{sn-bibliography}

\end{document}